%% file: bare_jrnl.tex
\documentclass[journal]{IEEEtran}

\usepackage{hyperref}

\ifCLASSINFOpdf
  \usepackage[pdftex]{graphicx}
  \graphicspath{{../Figures/}}
  \DeclareGraphicsExtensions{.pdf,.jpeg,.png}
\else
  \usepackage[dvips]{graphicx}
  \graphicspath{{../eps/}}
  \DeclareGraphicsExtensions{.eps}
\fi

\usepackage{amsmath}
\newcommand{\norm}[1]{\left\lVert#1\right\rVert}
\usepackage[ruled,vlined,shortend, linesnumbered]{algorithm2e} 
\usepackage[]{algorithmic} 
\usepackage{array}

\ifCLASSOPTIONcompsoc
  \usepackage[caption=false,font=normalsize,labelfont=sf,textfont=sf]{subfig}
\else
  \usepackage[caption=false,font=footnotesize]{subfig}
\fi

\usepackage{stfloats}

\usepackage{url}

\usepackage{algorithmic}

\usepackage{makecell,tabularx}

\usepackage{siunitx}

\usepackage[style=ieee, backend=biber, doi=false, maxnames=2, minnames=1]{biblatex}
\usepackage{xcolor}

\begin{document}
%
\title{Combined Transmission and Distribution State-Estimation for Future Electric Grids}
%
%
\author{Amritanshu Pandey,~\IEEEmembership{Member,~IEEE,}
        Shimiao Li,~\IEEEmembership{Student Member,~IEEE,}
        and~Larry~Pileggi,~\IEEEmembership{Fellow,~IEEE}
\thanks{All authors are affiliated with Electrical and Computer Engineering Department at Carnegie Mellon University.}
}

%
%

\markboth{Preprint version of the paper}%
{Shell \MakeLowercase{\textit{et al.}}: Bare Demo of IEEEtran.cls for IEEE Journals}


\maketitle

\begin{abstract}
\input{Abstract.tex}
\end{abstract}


\section{Introduction}
\input{Introduction.tex}

\input{PriorWork}

\input{Methodology}

\input{Experiments}

\section{Conclusion}

\input{Conclusion}

\subsection*{Reproducibility}
All data for the test cases is available in github repo: \url{git@gitlab.com:amritanshup/sugar_combinedtd_acse_testcases.git}

\printbibliography


%

\end{document}

%% file: Abstract.tex
Proliferation of grid resources on the distribution network along with the inability to forecast them accurately will render the existing methodology of grid operation and control untenable in the future. Instead, a more distributed yet coordinated approach for grid operation and control will emerge that models and analyzes the grid with a larger footprint and deeper hierarchy to unify control of disparate T\&D grid resources under a common framework. Such approach will require AC state-estimation (ACSE) of joint T\&D networks. Today, no practical method for realizing combined T\&D ACSE exists. This paper addresses that gap from circuit-theoretic perspective through realizing a combined T\&D ACSE solution methodology that is fast, convex and robust against bad-data. To address daunting challenges of problem size (million+ variables) and data-privacy, the approach is distributed both in memory and computing resources. To ensure timely convergence, the approach constructs a distributed circuit model for combined T\&D networks and utilizes node-tearing techniques for efficient parallelism. To demonstrate the efficacy of the approach, combined T\&D ACSE algorithm is run on large test networks that comprise of multiple T\&D feeders. The results reflect the accuracy of the estimates in terms of root mean-square error and algorithm scalability in terms of wall-clock time.

%% file: Introduction.tex
Today, the operation of the continental power grids is highly segregated with control of city-wide distribution grids isolated from control of nation-wide transmission grids. Such an approach is inherently sub-optimal, resulting in excessive use of carbon-intensive grid resources and marginalized reliability during natural disasters and other stressed scenarios. Tomorrow’s grid, therefore, will rely on an approach that enables joint operation and control of transmission and distribution (T\&D) resources. However, the present state of the art methods are inadequate for joint operation and control of T\&D grids since they were developed for grid sizes that span up to tens of thousands of transmission nodes whereas a real-time combined T\&D network would be represented by potentially hundreds of millions of nodes. The latter requires new models and methods that support re-envisioning the paradigm of grid operation to enable concurrent operation of both transmission and distribution grids in the most efficient, reliable, and resilient manner with inclusion of all grid resources.

Amongst the foremost and most fundamental analyses for joint T\&D grid operation is the combined T\&D AC state-estimation (ACSE) problem that would provide situational awareness for all T\&D resources and parameterize the T\&D network models for other analyses. No solution methodology exists today that can estimate the combined T\&D grid states robustly due to both the sheer size of the problem in terms of number of variables, as well as other practical requirements. Combined T\&D ACSE must accept sensor data from transmission and distribution grids as input and produce an estimate of grid states in near real-time. Disparate ACSE methods exist today for transmission \cite{Abur2004}, \cite{Monticelli2000}, \cite{Schweppe_SE} and distribution grids \cite{Baran_Node_Voltage_DSE} that are not generically extensible to estimation of combined T\&D network states. Furthermore, no standard method exists that can synchronously access and incorporate measurements from a wide array of sensors in the T\&D grids. This is partly due to the privacy needs of various T\&D stakeholders and partly due to the lack of a physics-based approach that can incorporate any type of sensor measurement into the grid model. The difficulty of the problem is further exacerbated by other challenges, such as inaccurate mapping of distribution grid topology and cyber-security concerns for devices ranging from critical grid assets to low-privilege access IoT devices.

To enable an envisioned scenario of joint control and operation of all T\&D resources, in this paper, we propose a new circuit-theoretic combined T\&D ACSE approach. By building it on a strong theoretic foundation, this approach enables distributed evaluation of the ACSE while supporting any existing or future T\&D grid components (including measurements) without loss of generality. We begin by \textit{distributed} circuit-theoretic construction of the ACSE problem by independently modeling different T\&D networks (with measurements included) as positive-sequence or three-phase equivalent circuits (see Section \ref{heading: Section III}). To couple these independent sub-network together, we develop a novel supervisory layer (see Section \ref{sec: supervisory layer}) to result in a combined T\&D ACSE model.  

Building upon the measurement-dependent combined T\&D circuit model, Section \ref{heading: Section V} formulates the circuit-theoretic ACSE problem. The formulation has two key properties: 1) convexity and 2) implicit robustness against bad data. To formulate a convex ACSE problem, Substitution Theorem \cite{ShimiaoLiSE} and feature mapping are applied when constructing the circuit models for measurement devices. The resulting circuit models are linear and impose affine network constraints. To ensure implicit robustness against bad-data, we build upon our prior work \cite{ShimiaoLiwlavSE}-\cite{ShimiaoLiSE} to utilize an inherent property of grid bad-data: that it is sparsely populated. Under the hold of this property, we minimize the weighted least absolute value (WLAV) of measurement error/noise to obtain a sparse residual vector that localizes suspicious bad-data. 

To make it scale well, as the number of T\&D grid states can be upwards of tens of millions, we develop a distributed ACSE computing strategy (regarding both memory and computing resources) away from traditional centralized approaches. To ensure timely solution of the complete problem, we apply theory of diakoptics (or node tearing) to breakdown the overall problem into sub-problems and we solve the overall problem without relaxation with Gauss-Seidel Newton (GSN) method. The efficacy and scalability of this approach is demonstrated in Section \ref{heading: Experiments}), by testing the distributed combined T\&D ACSE approach on large T\&D networks. To demonstrate the robustness of WLAV estimation under presence of bad-data, we compare the results from WLAV algorithm with a baseline of weighted least square (WLS) objective function.

%% file: PriorWork.tex
\section{Prior Work}
Combined T\&D ACSE is a subject that has not been studied extensively (exceptions include \cite{SUN_Global_SE}) even though traditional ACSE for transmission networks is a workhorse of day-to-day grid operation. The unique facets that make combined T\&D ACSE far more challenging than the traditional transmission or distribution ACSE are:
\begin{enumerate}
    \item dependency on a network model that is more complex and orders-of-magnitude larger than the traditional transmission or distribution network models, and;
    \item a need for a distributed estimation algorithm that is fast and resilient against bad or compromised measurement data, which are increasing in numbers due to growing attack surface.
\end{enumerate}
Hence, driven by the needs of these characteristics, the related prior work for combined T\&D ACSE can be categorized into two broad categories: i) approaches for combined modeling of T\&D networks and ii) approaches for distributed ACSE using these models.

\subsection{Combined T\&D Modeling}
The proposed approach for combined T\&D ACSE is dependent on the underlying combined T\&D models. There is active research on development of such combined T\&D models \cite{Kalsi_Integrated}, \cite{Gridspice}, \cite{Palmintier_TD}, \cite{master_slave_combined} \cite{Pandey_CombinedTD}; however, most of the research focus has been on building models for solving planning problems rather than state estimation that requires optimization and inclusion of measurement data.  These models, when combined, represent transmission and distribution grids with a mix of positive sequence and three-phase network components. Two modeling approaches stand out for such models: i) co-simulation approach \cite{Kalsi_Integrated}, \cite{Gridspice}, \cite{Palmintier_TD} and ii) co-modeling or integrated modeling approach \cite{master_slave_combined} \cite{Pandey_CombinedTD}.

In the co-simulation approach, separate tools model the transmission and distribution components with disparate formulations. Combined simulations are then run by communicating the coupling information between the tools. To achieve convergence, separate tools develop independent heuristics with a myopic view of its model rather than considering the complete T\&D context. Such an approach poses challenges for fast and timely convergence of the combined simulation as it is dependent on the success of each individual sub-simulation and the corresponding methodology.

Alternatively, co-modeling or integrated modeling approach enables a common framework for developing combined T\&D models. Literature for such frameworks is limited due to lack of common formulations for modeling both positive sequence transmission and three-phase distribution networks (\cite{master_slave_combined}, \cite{BackwardForwardSweep}, \cite{Current_Injection} and \cite{Pandey_Robust_Power}). One such approach by the authors overcomes this issue by modeling both transmission and distribution grid elements as circuit elements \cite{Pandey_CombinedTD}. Preliminary results based on these equivalent circuit models have demonstrated robust convergence for large T\&D networks for power flow problems \cite{Pandey_CombinedTD}. Here, we will extend the use of these circuit-theoretic models for combined T\&D ACSE problem with optimization capabilities.

The combined T\&D ACSE approach must incorporate measurement data from vast array of T\&D sensors. Although vast literature exists for incorporating standard metering devices on transmission networks (remote terminal units (RTUs) and phasor measurement units (PMUs) \cite{Schweppe_SE}, \cite{Hybrid_ISONE} and \cite{PMU_SE}), far fewer have discussed the incorporation of measurements on the distribution grid (\cite{DSE_Survey}, \cite{Branch_current_DSE} and \cite{Baran_Node_Voltage_DSE}) especially those consisting of new-age devices such as smart meters, EV charging etc. (\cite{SE_with_smart_meters} and \cite{Grijalva_AMI_Leveraging}). In Section IV, we will develop a generic methodology to include measurements from wide array of sensing devices from both transmission and distribution grids into the ACSE framework.

\subsection{Distributed State-Estimation Algorithms}

Even with the lack of substantive research in combined T\&D ACSE, there is plethora of active research in the field of distributed SE  \cite{Xie_Distributed_ACSE}, \cite{Minot_Distributed_SE_Transactions}, \cite{GUO_Distributed_SE}, \cite{Distributed_ACSE_Ross_ERCOT}) and distributed optimization for power grids  \cite{Distributed_Optimization_Survey_Molzahn}, \cite{Distributed_OPF_Microgrids}, \cite{Distributed_Optimization_Distribution}, \cite{Distributed_DCOPF_Kar_Hug} that target some facets of the described problem. However, most of this research is not directly applicable to solving combined T\&D ACSE due to its inability to efficiently scale to large combined T\&D networks with both positive sequence and three-phase components. A successful combined T\&D ACSE algorithm must be \textit{fast} and \textit{robust}. It should be significantly faster than the future T\&D market dispatch interval and it should provide certain convergence guarantees. Due to the implicit non-linearities in the traditional formulations \cite{Schweppe_SE}, \cite{Abur2004}, \cite{Monticelli2000}, these conditions cannot be satisfied.

\begin{figure}[htp]
    \centering
    \includegraphics[width=0.6\linewidth]{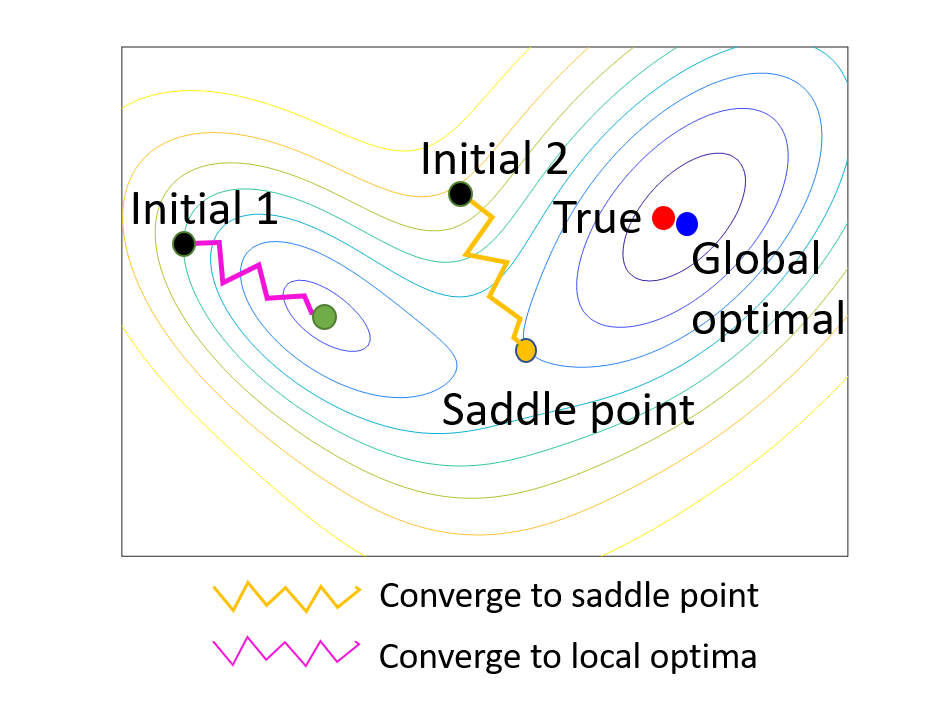}
    \caption{Contour plots for the solution space of weighted-least squares SE formulation.}
    \label{fig:contourWLSE}
\end{figure}

We consider the ubiquitous WLSE algorithm with power mismatch based measurement functions \cite{Abur2004}, \cite{Monticelli2000} to elaborate further. Due to the non-convex solution space (see the contour plot in Figure \ref{fig:contourWLSE}), the algorithm has multiple peaks and valleys that represent the various local optimal solutions for the ACSE problem (e.g. in the figure. \textcolor{green}{green} dot represents a local solution whereas \textcolor{blue}{blue} dot represents the global optimum). Depending on the choice of initial conditions and the corresponding convergence trajectory, the WLS formulation may converge to a local optimum or a global optimum. In case of convergence to a local optimum, we may observe high residuals with estimated states far from the true network states. Even worse it could converge to a saddle point (shown in \textcolor{yellow}{yellow}) again resulting in high differences between estimated and true network states. 

Some existing literature \cite{Hachtel_Constraints}, \cite{Linear_SE_Distribution_System} addresses these underlying challenges. For instance, many present provably convergent algorithms \cite{Linear_LAV_PMU}, \cite{Linear_SE_Distribution_System}; albeit almost all of them are only based on phasor measurement units (PMU) \cite{Linear_LAV_PMU}, \cite{PMU_SE} making them impractical in a realistic T\&D grid scenarios. There are methods that incorporate network topology to converge to a more physical solution, but they are known to face numerical difficulties \cite{Hachtel_Constraints} with added convergence challenges. There is also recent works on hybrid approach \cite{Hybrid_ISONE} that include both RTU and PMU measurements into its framework; however, these are still formulated as a non-convex optimization problem, thereby facing the same challenges as those described above. More recently, the authors have proposed centralized circuit-theoretic WLSE and WLAV approaches for transmission grid ACSE \cite{ShimiaoLiSE}. These tackle some challenges with traditional transmission ACSE formulation but does not directly extend to distribution grids or distributed computing frameworks.

%% file: Methodology.tex
\section{Equivalent Circuit Modeling for Combined T\&D Networks} \label{heading: Section III}

\begin{figure}[htp]
    \centering
    \includegraphics[width=0.9\linewidth]{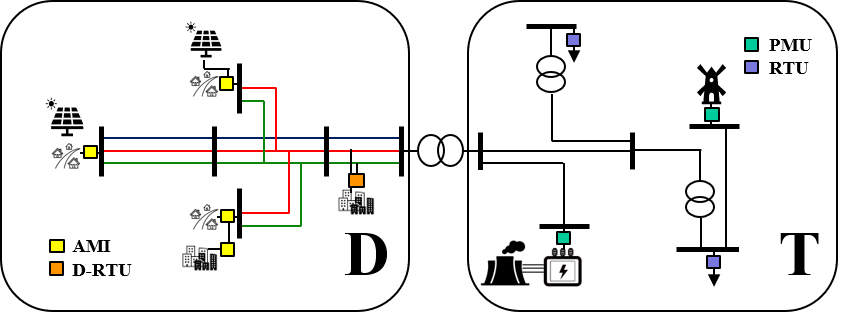}
    \caption{Notional figure of combined T\&D network.}
    \label{fig:TD_cartoon_equip}
\end{figure}
\begin{figure}[htp]
    \centering
    \includegraphics[width=0.9\linewidth]{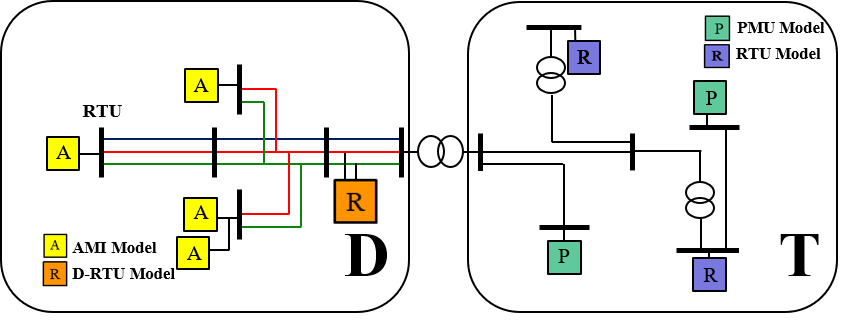}
    \caption{Notional figure of combined T\&D network with measurement model: measurement components have been replaced by circuit models so that the coupled T\&D network is equivalent to interconnected sub-circuits.}
    \label{fig:TD_cartoon_meas}
\end{figure}

\noindent The combined T\&D ACSE framework developed within this paper is based on a combined T\&D model with its data distributed across multiple memory units and ownership across multiple stakeholders. We construct the model following a circuit-theoretic approach \cite{Pandey_CombinedTD} that represents transmission network components as positive sequence circuits and distribution network components as three-phase circuits. To couple the various sub-circuit models ($S \subseteq \textbf{S}$), we develop a supervisory layer (preferably on cloud) that communicates between various sub-network entities through coupling circuits to enable a robust distributed optimization framework. We apply Substitution Theorem to map measurement data into circuit models via replacing any measured element in the combined T\&D grid (see Figure \ref{fig:TD_cartoon_equip}) with a mix of independent voltage and current sources (see Figure  \ref{fig:TD_cartoon_meas}). To model measurement noise, we augment the sources with slack injections to satisfy Kirchhoff's current laws. The aggregated measurement-dependent circuit includes complete topology of the T\&D network, including the zero-injection nodes which constitute a majority of the distribution system. This constricts the solution space of the estimator and drives it to a more physically meaningful solution.

\subsection{Measurement Models for Combined T\&D Networks}
The electric grid consists of a variety of sensing equipment measuring various system states at different sampling rates. Most of these measurements are communicated back to grid T\&D control rooms owned by entities such as local grid operators, utilities and data aggregators. This section describes the mapping of measurement data into measurement circuit models, which are stored at the local entity level. Even though variety of sensing devices exist in the T\&D grid (e.g., RTUs, smart meters, IoTs etc.), all measured parameters of interest from these devices can be categorized into one of the following four categories without loss of generality:
\begin{itemize}
    \item Root mean square (RMS) power injection and flow measurements
    \item RMS voltage nodal measurements
    \item Phasor voltage nodal measurements
    \item Phasor current injection and flow measurements
\end{itemize}
For the transmission network each measurement parameter represents a positive sequence quantity whereas for distribution network, it represents a three-phase quantity. Starting from fundamental block of these four categories, we can construct the measurement circuit model for any grid measuring device. We describe two examples in detail: (i) transmission grid PMU and (ii) distribution grid smart meter. Following the same approach, other circuit models for other measurement devices can be constructed.
 
\subsubsection{Phasor measurement units on the transmission network}

\begin{figure}[htp]
    \centering
    \includegraphics[width=\linewidth]{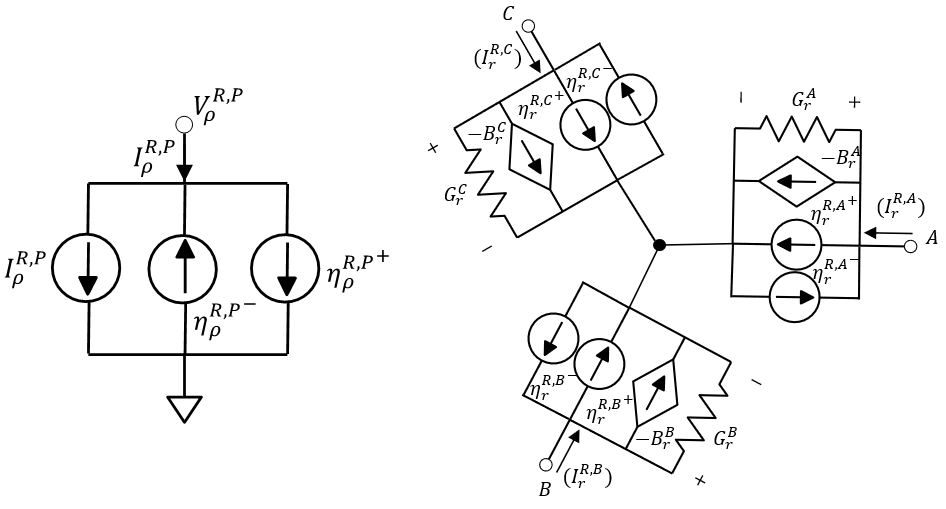}
    \caption{Real part of PMU injection circuit model (left); Real part of a smart-meter circuit model for wye connected load. (right).}
    \label{fig:pmu_injection}
\end{figure}
 
Phasor measurement units (PMUs) are highly accurate measurement devices with clock synchronization through a common GPS signal. These devices can capture voltage and current waveforms concurrently with other such PMU devices. This allows a PMU device (identified by $\rho$ in subscript of its parameters) to measure voltage and current magnitudes as well as the corresponding angle differences with respect to a system reference. Injection and flow measurements constitute two critical measurements by a PMU device. To further demonstrate mapping of a PMU device to an equivalent circuit model, we make use of injection measurements as an example.
 
For injection measurements, a PMU can provide real-time meter readings of positive sequence $(P)$ voltage and injection current phasors: $(I_{\rho}^{R, P}, I_{\rho}^{I, P}, V_{\rho}^{R, P}, V_{\rho}^{I, P})$ located at a set of network nodes $N_{\rho} \subseteq N$. By substitution theorem, we map these PMU measurements into an equivalent circuit model (shown in Figure \ref{fig:pmu_injection} (left)). The model is represented by independent current sources with values corresponding to measured real and imaginary currents. To further include measurement noise such that the KCL is always satisfied, we add additional slack current sources $\eta$ to the model such that:
\begin{alignat}{3}  
    I_{i}^{R,P} = I_{\rho,i}^{R, P} + {\eta_{\rho,i}^{R, P}}, & \quad \forall i \in N_{\rho}\label{eq:PMU real, single noise term}\\
    I_{i}^{I,P} = I_{\rho,i}^{I, P} + {\eta_{\rho,i}^{I,P}}, & \quad \forall i \in N_{\rho}
    \label{eq:PMU imag, single noise term}
\end{alignat}

To provide robustness against bad data, this paper builds a WLAV method that minimizes least absolute value of measurement error, i.e., solving a solution that satisfies $\min ||\eta||_1$. To convert this L1-norm term into an equivalent but differentiable form, we split the noise term $\eta$ into 2 non-negative variables  
${\eta^{+},\eta^{-}}$, such that $\eta=\eta^+-\eta^-,
||\eta||_1=\eta^++\eta^-,$ and $\eta^+,\eta^-\succeq0$. Thus the model in (\ref{eq:PMU real, single noise term})-(\ref{eq:PMU imag, single noise term}) is equivalently represented as:
\begin{alignat}{3}  
    I_{i}^{R,P} = I_{\rho,i}^{R, P} + {\eta_{\rho}^{R,P}}^{+} - {\eta_{\rho}^{R,P}}^{-}, & \quad \forall i \in N_{\rho}\\
    I_{i}^{I,P} = I_{\rho,i}^{I, P} + {\eta_{\rho}^{I,P}}^{+} - {\eta_{\rho}^{I,P}}^{-}, & \quad \forall i \in N_{\rho}
\end{alignat}

\noindent 
Models in the rest of this paper are also represented with $\eta^+,\eta^-$ variables to fit in the WLAV-based problem definition.

Even though the measured current values can completely parameterize the states of the measured node, PMUs also provide redundant measurements in terms of voltage phasors $(V_{\rho}^{R, P}, V_{\rho}^{I, P})$. We include these measurements into the formulation for improving estimation quality as additional constraints (see eqns. \eqref{eq:problem_definition}, \eqref{eq:voltage_constraint_PMU}, \eqref{eq:problem_definitiong}). Following the same methodology, positive sequence and three-phase models for PMU flow measurements can also be constructed.

\subsubsection{Smart meter for three-phase wye connected load}

 
Smart meters or AMI infrastructure are becoming the most ubiquitous sensing devices on the distribution grid. Although they are rarely used for grid situational awareness today, in the near future, this is likely to change. These smart meters (identified by $r$ subscript in its parameters) measure three-phase complex voltage magnitudes along with real and reactive power quantities; however, unlike PMU or $\mu$PMUs devices they cannot measure relative angles for the voltage or current quantities. Depending on the connection type, these meters can measure single, triplex or three-phase voltages and powers. As an example, we show construction of an equivalent circuit model for a smart meter that is wired across a wye connected three-phase load. The meter for each phase $(\delta \in {A, B, C})$ measures nodal voltage ($|V_r^{\delta}|$) and power quantities ($P_r^{\delta}$ and $Q_r^{\delta}$) on the network nodes $N_r \subseteq N$. However, these measured parameters are noisy and therefore the \textit{true} parameters for each phase $\delta$ are represented by:
\begin{alignat}{3}
    P_i^{\delta} = P_{r, i}^{\delta} + \Delta P_{r, i}^{\delta}, & \quad \forall i \in N_r \\
    Q_i^{\delta} = Q_{r, i}^{\delta} + \Delta Q_{r, i}^{\delta}, & \quad \forall i \in N_r \\
    |V_i^{\delta}| = |V_{r, i}^{\delta}| + \Delta |V_{r, i}^{\delta}|, & \quad \forall i \in N_r
\end{alignat}

\noindent To represent the measured parameters as an equivalent circuit, we perform a feature transformation:
\begin{alignat}{3}
    G_{r,i}^{\delta} = \frac{P_{r,i}^{\delta}}{{|V_{r,i}^{\delta}|}^2}, & \quad \forall i \in N_r\\
    B_{r,i}^{\delta} = \frac{Q_{r,i}^{\delta}}{{|V_{r,i}^{\delta}|}^2}, & \quad \forall i \in N_r
\end{alignat}

\noindent and to capture noise elements in the model to satisfy KCL constraints, we add slack injections ($\eta_{r}^{ \delta}$). As in the case of PMU model, we separate the noise terms into positive and negative components to facilitate WLAV estimation:
\begin{alignat}{3}
    I_{r, i}^{R, \delta} = G_{r, i}^{\delta} V_i^{R, \delta} - B_{r, i}^{\delta} V_i^{I, \delta} + {\eta_{r, i}^{R, \delta}}^+ - {\eta_{r, i}^{R, \delta}}^-,& \quad \forall i \in N_r \label{eq:8}\\ 
    I_{r, i}^{I, \delta} = G_{r, i}^{\delta} V_i^{I, \delta} + B_{r, i}^{\delta} V_i^{R, \delta} + {\eta_{r, i}^{I, \delta}}^+ - {\eta_{r, i}^{I, \delta}}^-,& \quad \forall i \in N_r \label{eq:9}
\end{alignat}

\noindent To construct a measurement device model, \eqref{eq:8}-\eqref{eq:9} are mapped into an equivalent circuit which then replaces the wye-connected load that it senses (see Figure \ref{fig:TD_cartoon_meas}). Figure \ref{fig:pmu_injection} (right) represents the real circuit for the smart meter measuring three-phase wye connected load.

The described modeling approach for smart meter does a feature transformation that reduces three measured parameters $(|V_{r, i}|, P_{r, i}, Q_{r, i})$ to two $(G_{r,i}, B_{r,i})$. This could result in loss of critical information. Therefore, we can further minimize the L1-norm of square of voltage magnitude noise $(|V_{r, i}|^2 - {V^R_i}^2 - {V^I_i}^2|)$ in the objective to preserve this information; however, doing so nullifies the convexity of the overall formulation.

\subsubsection{Assumptions on model observability}
Global observability of T\&D network is a necessary condition for running a combined T\&D ACSE. To satisfy this condition, the developed T\&D model must satisfy the following:\\
\textit{The system matrix $(Y_S)$ of each sub-network $(S \subseteq \textbf{S})$ that represents nodal balance equations at all studied nodes must be full-ranked (i.e. an inverse ${(Y_S)}^{-1}$ exists).}

More generally stated, this implies whether the set of flow and injection measurements in the sub-network are sufficient to describe all the states in the grid. Transmission networks not only have sufficient measurements to describe the state of the network, but they generally have redundant measurements to further improve the quality of estimates \cite{Schweppe_SE}. We include these redundant measurements into the framework (see \eqref{eq:problem_definition}) as constraints.
However, ensuring observability for distribution networks remains an active research area \cite{Distribution_Grid_Observability}, \cite{DSE_Parameterization_Observability}. Here we assume that we have access to measurements at all non zero-injection nodes in the studied distribution grids. These can include high accuracy substation-based intelligent electronic devices (IEDs) and $\mu$PMU devices, as well as smart meters or newer cheap IoT devices. Nonetheless, there is a potential to explore techniques such as pruning of networks (topology simplification or reduction), use of forecast as measurements, use of low quality AMI data and other methods such matrix completion \cite{Donti_matrix_completion} for improving combined T\&D model observability. However, this is not covered within the scope of this paper alongside discussion of techniques for time-synchronization of this data.

\section{Supervisory Layer for Combined T\&D ACSE}\label{sec: supervisory layer}

Section \ref{heading: Section III} discussed the construction of measurement-dependent circuit models for distinct T\&D utilities. In this section, we describe how these T\&D disjoint models are coupled to create a unified combined T\&D circuit model. To couple various distributed T\&D models, we construct a supervisory layer that is presumably hosted on the cloud and communicates information about boundary nodes between various distinct utilities. In \cite{Pandey_CombinedTD}, we developed a similar supervisory layer for combined T\&D power flow. Here, the approach is extended to enable online distributed optimization in presence of measurements. To capture knowledge of boundary nodes, supervisory layer consists of a set of coupling circuits, which are an aggregation of controlled current and voltage sources. These circuits couple (or unify) the otherwise partitioned circuit models.

\begin{figure}[t]
    \centering
    \includegraphics[width=4.5cm]{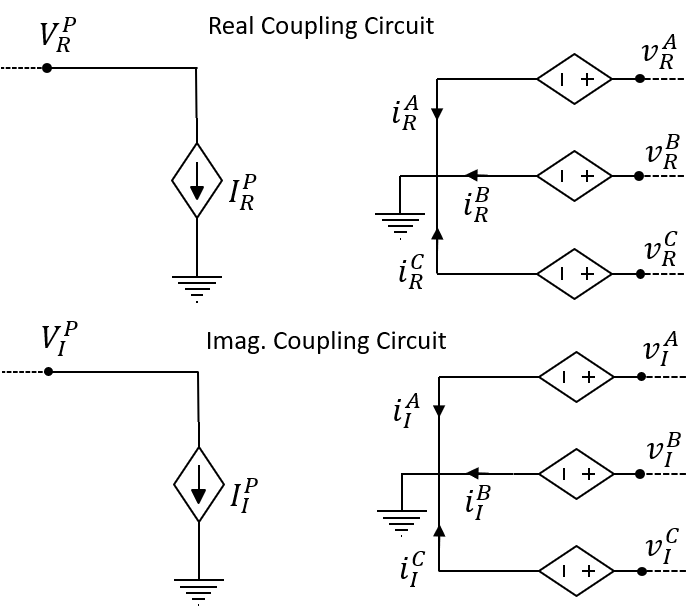}
    \caption{Primal coupling circuit between T\&D network models.}
    \label{fig:coupling circuit}
\end{figure}

Figure \ref{fig:coupling circuit} depicts the equivalent circuit for the coupling circuit between positive sequence T and three-phase D circuits. The set of nodes on left of the figure represent the nodes for the coupling circuit on the transmission side for bus $m$, whereas the set of nodes on the right of the figure represent the nodes on the distribution end for three-phase bus $m$. As shown in Figure \ref{fig:coupling circuit}, the controlled current sources $(I_R^P$ and $I_I^P)$ and controlled voltage sources $(v_R^A , v_R^B , v_R^C , v_I^A , v_I^B, v_I^C)$ represent the coupling variables between the transmission and distribution sub-circuits. Specifically, the controlled current sources $(I_R^P$ and $I_I^P)$ represent the positive sequence currents to be delivered to the distribution sub-circuit, whereas controlled voltage sources represent the nodal voltages for the distribution grid $(v^{ABC})$ that mirrors the positive sequence transmission voltages $(V^{P})$ at the point of interconnection (POI). Importantly, this port enables connection of any number of transmission and distribution networks without loss of generality and together they include the set of nodes ${N}_c^P$ and ${N}_c^{ABC}$ to represent all nodes for coupling circuits at transmission and distribution side, respectively. 
The theory of symmetrical components is used to derive the expressions for coupling currents and voltages required to model the port. The positive sequence model of the power grid assumes balanced operation of the grid, and therefore, ignores the zero and negative sequence components of voltages and currents. Hence, by ignoring the zero and negative sequence currents consumed by the distribution feeder at POI, we first model the current coupling between T\&D (as depicted in Figure \ref{fig:coupling circuit}) through the relationship characterized by controlled current sources, see (\ref{eq: T&D current coupling}) where $\alpha$ denotes the value of $\frac{2\pi}{3}$ radians:
\begin{equation}
\begin{bmatrix}
I_R^P\\I_I^P
\end{bmatrix}
=
\frac{1}{3}\begin{bmatrix}
1&0&\alpha&0&\alpha^2&0
\\0&1&0&\alpha&0&\alpha^2
\end{bmatrix}
\begin{bmatrix}
i_R^A\\i_I^A\\i_R^B\\i_I^B\\i_R^C\\i_I^C
\end{bmatrix}
\label{eq: T&D current coupling}
\end{equation}

Similarly, the voltage coupling between transmission and distribution (also depicted in Figure \ref{fig:coupling circuit}) is modeled by formulating the distribution end voltages as a function of transmission POI voltages, as (\ref{eq: T&D voltage coupling}) shows:

\begin{equation}
\begin{bmatrix}
v_R^A\\v_I^A\\v_R^B\\v_I^B\\v_R^C\\v_I^C
\end{bmatrix}
=
\begin{bmatrix}
1&0\\
0&1\\
\alpha^2&0\\
0&\alpha^2\\
\alpha&0\\
0&\alpha\\
\end{bmatrix}
\begin{bmatrix}
V_R^P\\V_I^P
\end{bmatrix}
\label{eq: T&D voltage coupling}
\end{equation}

The coupling circuits above describe the relationship between primal states of T\&D boundary nodes. However, for combined T\&D ACSE (i.e. distributed optimization problem), we derive similar relationships for dual optimization variables as well (See Section \ref{heading: Section V}). In this approach, we ensure that minimum data has to communicated from each utility to the supervisory layer. So far it only includes states corresponding to the boundary nodes. Additionally, if zero injection nodes are chosen as boundary nodes, then various utilities do not have to share any measurement data with the supervisory layer but rather only the solutions of voltages and currents at the boundary node. This helps satisfy the privacy needs of the various utilities. Next we describe the distributed combined T\&D ACSE algorithm that actuates upon this unified circuit model.

\section{Distributed Combined T\&D weighted-LAV ACSE Formulation}\label{heading: Section V}

The general idea behind the proposed combined T\&D ACSE approach is to minimize the norm of slack injections in measurement devices subject to the Kirchhoff's Current Laws (KCL) at each node in the unified circuit model. Depending on the chosen norm, the approach can be described as weighted least square estimation (WLSE) or weighted least absolute value (WLAV) estimation problem. While WLSE algorithm is more prominent due to its equivalence to maximum likelihood estimation (MLE) under presence of Gaussian noise, it is not robust against bad-data. As many distribution grid measurements are likely to be erroneous, a robust estimator like the WLAV is preferred, which is statistically equivalent to MLE when noise is modeled as Laplacian distribution. 

The solution approach inputs measurement and network data in the form of a distributed T\&D circuit model. It outputs robust estimates for the T\&D grid states. The network constraints are affine due to the use of circuit-theoretic modeling approach (see Section \ref{heading: Section III} and \cite{ShimiaoLiwlavSE} for details). The resultant optimization problem is convex with a linear objective, affine equality constraints and convex inequality constraints. The convexity of the formulation in addition to the natural weak coupling between T\&D networks allows efficient use of distributed algorithms for solution of the problem. This helps ensure timely convergence of the combined T\&D ACSE algorithm with provable guarantees of convergence; necessary conditions to be of practical use. The formulation is mathematically described in \eqref{eq:problem_definition}. Note that inequality constraints exist in addition to KCL at all nodes due to transformation of non-differentiable L1 terms of the WLAV objective to mathematically equivalent continuous twice-differential terms through separation of each noise term into two variables $(+/-)$. Having this form allows use of the distributed primal-dual interior point (PDIP) algorithm \eqref{eq:problem_definition} with the Newton's method.

\begin{subequations}\label{eq:problem_definition}
\begin{small}
\begin{alignat}{3}
    \min_{\eta, V} \quad & \sum_{i \in N_{\rho}} w_{\rho, i}\left({\eta_{\rho, i}^{R, \delta}}^+ + {\eta_{\rho, i}^{R, \delta}}^- + {\eta_{\rho, i}^{I, \delta}}^+ + {\eta_{\rho, i}^{I, \delta}}^-\right) \notag\\
    \quad &+ \sum_{i \in N_{\rho}}w_{\rho, i}\left({\eta_{\rho, i}^{VR, \delta}}^+ + {\eta_{\rho, i}^{VR, \delta}}^- + {\eta_{\rho, i}^{VI, \delta}}^+ + {\eta_{\rho, i}^{VI, \delta}}^- \right)\notag\\
    \quad & + \sum_{i \in N_{r}} w_{r, i}\left({\eta_{r, i}^{R, \delta}}^+ + {\eta_{r, i}^{R, \delta}}^- + {\eta_{r, i}^{I, \delta}}^+ + {\eta_{r, i}^{I, \delta}}^-\right) \notag\\
    \quad & + \sum_{ij \in N_{\rho}} w_{\rho, ij}\left({\eta_{\rho, ij}^{R, \delta}}^+ + {\eta_{\rho, ij}^{R, \delta}}^- + {\eta_{\rho, ij}^{I, \delta}}^+ + {\eta_{\rho, ij}^{I, \delta}}^-\right) \notag\\
    \quad & + \sum_{ij \in N_{r}} w_{r, ij}\left({\eta_{r, ij}^{R, \delta}}^+ + {\eta_{r, ij}^{R, \delta}}^- + {\eta_{r, ij}^{I, \delta}}^+ + {\eta_{r, ij}^{I, \delta}}^-\right)
\end{alignat}
\begin{flalign}
& \textrm{subject to:} & \notag \\
& \textrm{\underline{Nodal constraints for current injection phasor measurements}} & \notag
\end{flalign}
\begin{alignat}{3}
    V^T Y_i^R + I_{\rho, i}^{R, \delta} + {\eta_{\rho, i}^{R, \delta}}^+ - {\eta_{\rho, i}^{R, \delta}}^- = 0, & \quad \forall i \in N_{\rho}\\
    V^T Y_i^I + I_{\rho, i}^{I, \delta} + {\eta_{\rho, i}^{I, \delta}}^+ - {\eta_{\rho, i}^{I, \delta}}^- = 0, & \quad \forall i \in N_{\rho}
\end{alignat}
\begin{flalign}
& \textrm{\underline{Nodal constraints for RMS power injection measurements}} & \notag
\end{flalign}
\begin{alignat}{3}
    V^T Y_i^R + G_{r, i}^{\delta} V_i^{R, \delta} - B_{r, i}^{\delta} V_i^{I, \delta} + {\eta_{r, i}^{R, \delta}}^+ - {\eta_{r, i}^{R, \delta}}^- = 0,& \quad \forall i \in N_r\\
    V^T Y_i^I + G_{r, i}^{\delta} V_i^{I, \delta} + B_{r, i}^{\delta} V_i^{R, \delta} + {\eta_{r, i}^{I, \delta}}^+ - {\eta_{r, i}^{I, \delta}}^- = 0,& \quad \forall i \in N_r
\end{alignat}
\begin{flalign}
& \textrm{\underline{Constraints for phasor voltages}} & \notag
\end{flalign}
\begin{alignat}{3}
    V_{i}^{R, \delta} - V_{\rho,i}^{R, \delta}+ {\eta_{r, i}^{VR, \delta}}^+ - {\eta_{r, i}^{VR, \delta}}^-=0, & \quad \forall i \in N_{\rho}\label{eq:voltage_constraint_PMU}\\
    V_{i}^{I, \delta} - V_{\rho,i}^{I, \delta}+ {\eta_{r, i}^{VI, \delta}}^+ - {\eta_{r, i}^{VI, \delta}}^-=0, & \quad \forall i \in N_{\rho}\label{eq:problem_definitiong}
\end{alignat}
\begin{flalign}
& \textrm{\underline{Nodal constraints for zero-injection (ZI) nodes}} & \notag
\end{flalign}
\begin{alignat}{3}
    V^T Y_{i}^R = 0, & \quad  \forall i \in N_{ZI}\\
    V^T Y_{i}^I = 0, & \quad  \forall i \in N_{ZI}
\end{alignat}
\begin{flalign}
& \textrm{\underline{Nodal constraints for RMS flow measurements}} & \notag
\end{flalign}
\begin{alignat}{3}
    I_{ij}^{R, \delta} 
    + G_{r,ij}^{\delta} V_i^{R, \delta}
    - B_{r,ij}^{\delta} V_i^{I, \delta} + {\eta_{r, ij}^{R, \delta}}^+ - {\eta_{r, ij}^{R, \delta}}^- = 0 
    \\
    I_{ij}^{I, \delta}
    + G_{r,ij}^{\delta} V_i^{I, \delta}
    + B_{r,ij}^{\delta} V_i^{R, \delta} + {\eta_{r, ij}^{I, \delta}}^+ - {\eta_{r, ij}^{I, fl, \delta}}^- = 0\\ 
    ij \in E_r  \subseteq{\{(i,j) \quad | \quad i \in N_r, j \in N},j\neq i\}\notag
\end{alignat}
\begin{flalign}
& \textrm{\underline{Nodal constraints for phasor flow measurements}} & \notag
\end{flalign}
\begin{alignat}{3}
    I_{\rho,ij}^{R, \delta} - I_{ij}^{R, \delta} + {\eta_{\rho, i}^{R, \delta}}^+ - {\eta_{\rho, i}^{R, \delta}}^- = 0 
    \\
    I_{\rho,ij}^{I, \delta} - I_{ij}^{I, \delta} + {\eta_{\rho, i}^{I, \delta}}^+ - {\eta_{\rho, i}^{I, \delta}}^- = 0\\ ij \in E_\rho  \subseteq{\{(i,j) \quad | \quad i \in N_{\rho}, j \in N}, j\neq i \}\notag
\end{alignat}
\begin{flalign}
& \textrm{\underline{Positivity of all noise terms}} & \notag
\end{flalign}
\begin{alignat}{3}
    \eta^{+/-}\succeq 0
\end{alignat}
\end{small}
\end{subequations}
In \eqref{eq:problem_definition}, $N_{\rho}$ is a subset of network nodes $N$ to which phasor measurement devices are connected $(N_{\rho} \subseteq N)$ and $N_r$ is a subset of network nodes $N$ to which RMS measurement devices are connected $({N_r \subseteq N})$, $E_{\rho}$ and $E_r$ denotes the set of edges (branches) connected to the phasor and RMS measurement devices. $w$ is the weight associated to trustworthiness of each measurement and is a function of device type and tolerance. ${\delta}$ is the set of phases that for three-phase networks represent phases: $\{A, B, C\}$ and for transmission networks represent the positive sequence component: $\{P\}$. To solve the problem, we implement a distributed PDIP algorithm augmented with circuit-theoretic limiting methods \cite{Pandey_Robust_Power} and \cite{tx_stepping} to ensure timely convergence. We describe the algorithm in Algorithm \ref{alg:framework}. In contrast to general purpose optimization tools where the entire vectors of primal and dual optimization variables are limited in the PDIP approach \cite{IPOPT}, \cite{Coleman1996}, we individually damp the primal and dual variables based on their physical characteristics. For instance, grid-physics indicates that a Newton step for the voltage variable of \textit{2 pu} exhibits non-physical behavior and therefore we significantly damp it. Similarly for ensuring dual feasibility, we apply limiting on the $\mu$ dual variables (see \cite{Jereminov2019_OPF} for details). Following such methodology in the past, we have demonstrated robust convergence of power flow and AC-optimal power flow problems \cite{Pandey_Robust_Power}, \cite{Jereminov2019_OPF}. Here we apply these heuristics to the circuit based distributed WLAV-ACSE algorithm. 

\begin{figure}[t]
  \centering
  \includegraphics[width=1.1\linewidth]{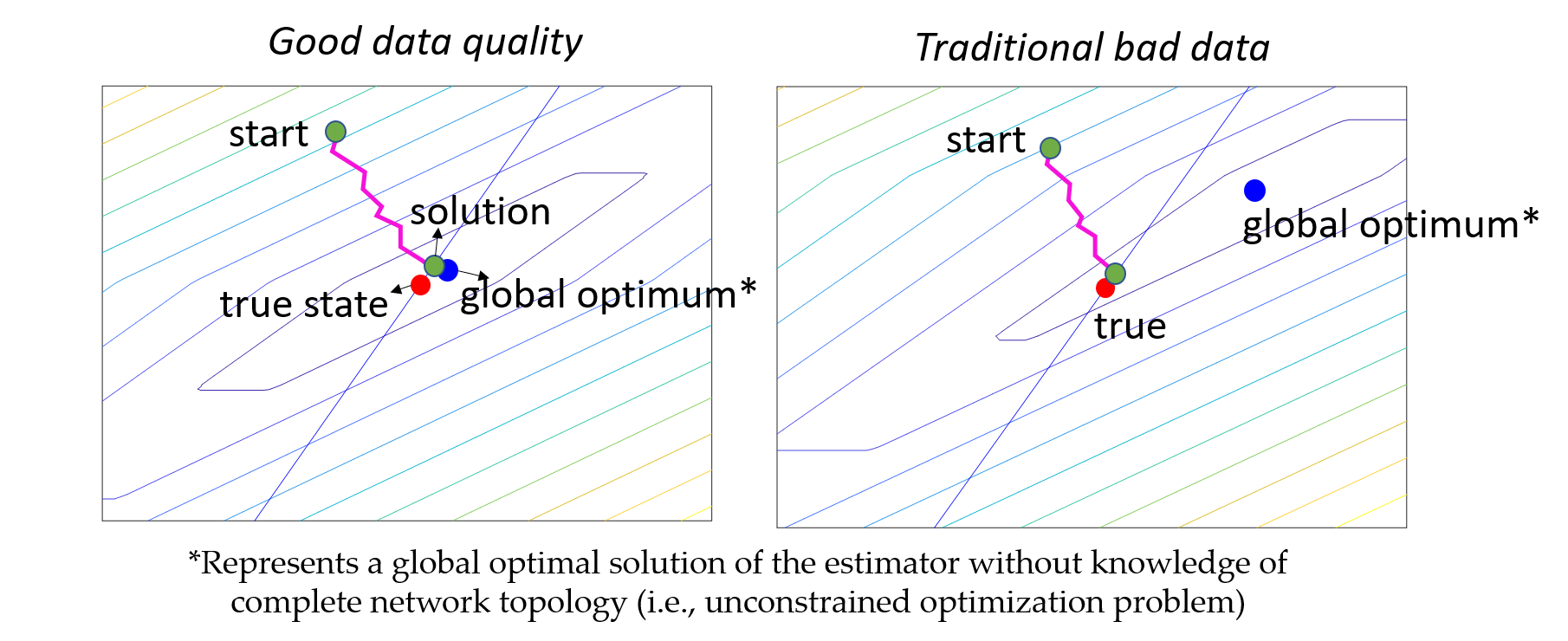}
  \caption{Contour plots of the solution space for WLAV-ACSE with and without bad-data.}
  \label{fig:LAV_SE}
\end{figure}

The circuit-theoretic distributed WLAV-ACSE formulation in \eqref{eq:problem_definition} has significant advantages over the traditional PQV-based WLSE/WLAV formulations. Traditional PQV formulations are highly non-convex (also see Figure \ref{fig:contourWLSE}) and are unable to scale to robust distributed frameworks. The non-convexity also makes it difficult to obtain timely solution for large combined T\&D ACSE problems. In contrast, the developed circuit-theoretic formulation is convex with only one optimum solution (see Figure \ref{fig:LAV_SE}) and therefore can provide fast convergence guarantees for the grid-physics constrained optimization problem.  In addition, the proposed WLAV approach is robust against bad-data samples  and can converge to a solution closer to \textit{true} state even in presence of many bad-data samples (see right of Figure \ref{fig:LAV_SE} and \cite{ShimiaoLiwlavSE}).

\section{Distributed Computing Framework for Combined T\&D WLAV ACSE Problem}

To solve the problem described in \eqref{eq:problem_definition}, we develop a distributed primal dual interior point approach that can operate on scattered data spread across multiple utilities. This \textit{distributed} approach is mandated by practical realities of T\&D grid problems:

\begin{itemize}
    \item Stakeholders representing separate T\&D utilities may not intend to share all internal network data with each other.
    \item Solving an iteration matrix with hundreds of millions of nodes may not be viable under a single machine single memory computing scheme.
\end{itemize}

The inherent \textit{weak} coupling between the T\&D sub-networks (i.e., most instances of T\&D coupling assumes radial connection) renders distributed solution algorithm apt for this problem while satisfying the necessary provisions of \textit{speed, robustness,} and \textit{resilience}. More specifically, it enables domain-based decomposition (\cite{Kron}, \cite{Newton_Vincentelli_Relaxation} and \cite{Vlach1985_Decomposition}) of the overall problem through node-tearing such that the solution of the overall problem is obtained via concurrent runs of decomposed sub-problems.

\begin{figure}[t]
    \centering
    \includegraphics[width=8cm]{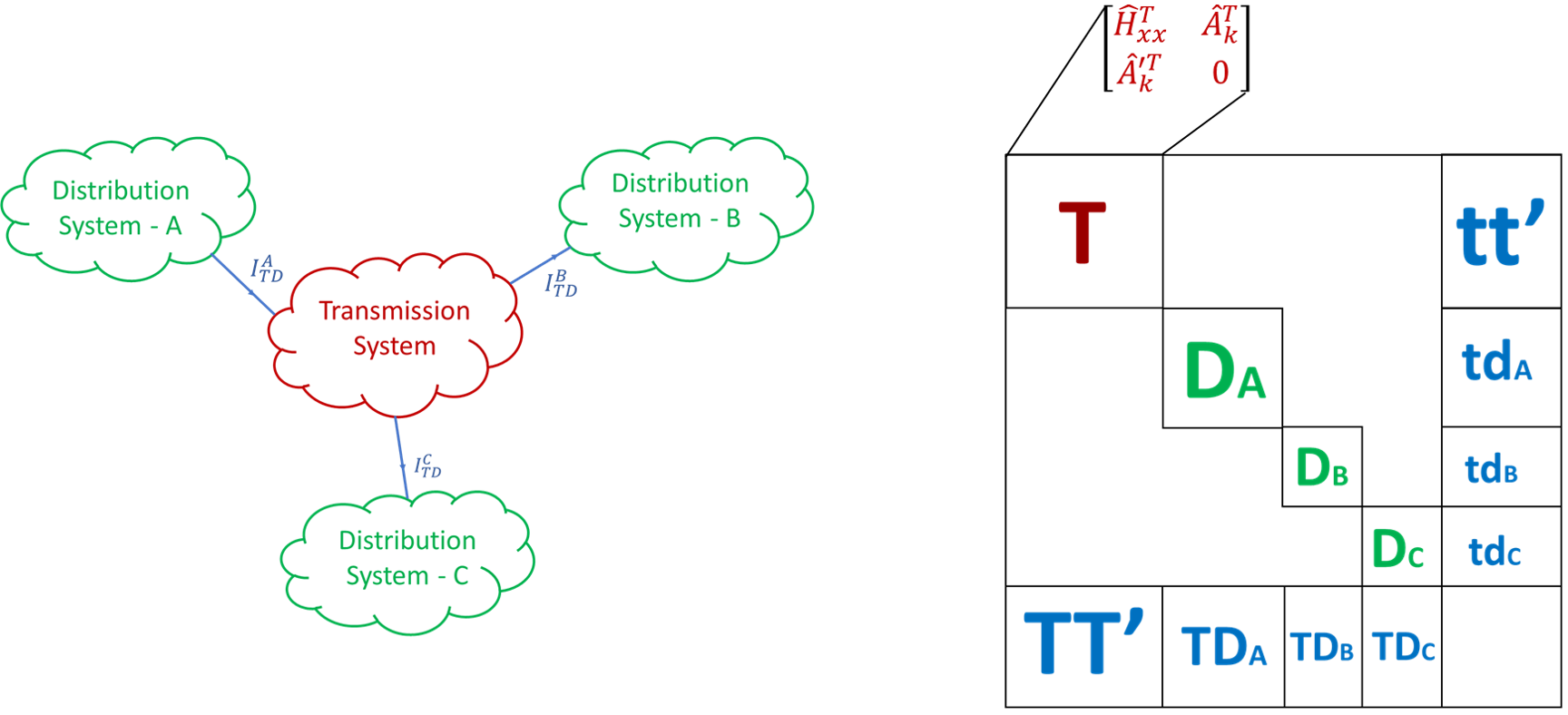}
    \caption{Combined T\&D ACSE solution matrix in BBD structure, where $H_{xx}=\nabla_X^2f_{obj},A=\nabla_Xg(X)$.}
    \label{fig:BBD_matrix}
\end{figure}

To describe the solution methodology, we first symbolically derive the perturbed first-order optimality conditions (KKT) in \eqref{eq:perturbed} for the problem described in \eqref{eq:problem_definition}. The solution to the perturbed KKT equations in \eqref{eq:perturbed}, for a convex problem described in \eqref{eq:problem_definition}, provides the WLAV estimate $X^*$ of the combined T\&D ACSE problem as well as the global optimum of the underlying optimization problem. However, for combined T\&D ACSE the set of equations can be very large and data scattered across many utilities. Therefore, instead of solving the perturbed KKT condition centrally, we decompose them into many sub-problems through node-tearing. To do so, we first represent it in the bordered-block diagonal form as shown in Figure \ref{fig:BBD_matrix}.
\begin{subequations}\label{eq:perturbed}
\begin{alignat}{3}
\nabla_\lambda\mathcal{L}=g\left(X\right)=0\\
\nabla_X\mathcal{L}=\nabla_X f_{obj}(X)+\nabla_X^T g\left(X\right)\mathit{\lambda}-\mathit{\mu}=0\\
-\mathit{\mu}\ \odot\left(X-\underline{X}\right)+\epsilon=0\\
\mathit{\mu}\geq\mathbf{0}\\
X\geq\underline{X}
\end{alignat}
\end{subequations}

\noindent where, $X\in \{V^{R,I}, \eta^{R,I}\}$ represent the set of system states, $f_{obj}$ denotes the objective function, $g(X)$ denotes the set of equality constraints, 
$X\geq\underline{X}$ represents the lower bound constraints for the $\eta^{+/-}$ variables, $\lambda$ is the set of Lagrangian multipliers of equality constraints, and $\mu$ denotes the Lagrangian multipliers for inequality constraints.

To apply the distributed approach to \eqref{eq:perturbed}, we divide the solution variables $X$ into $|\textbf{S}|$ sets of $X^S_{int}\in \{V_{int, S}^{R,I}, \eta_S^{R,I}\}$ and $X^S_{ext}\in \{V_{ext, S}^{R,I}\}$, where $S\in\textbf{S}$ and $\textbf{S}$ is the set of all the subsystems. Similarly, the set of dual variables are separated into distinct sets as well, with $\lambda$ separated into $\lambda^S_{int}\in \{\lambda_{int, S}^{R,I}\}$, $\lambda^S_{ext}\in \{\lambda_{ext, S}^{R,I}\}$, and $\mu$ are separated into $|\textbf{S}|$ sets of $\mu_S$. The overall problem \eqref{eq:perturbed} is decomposed into $|\textbf{S}|$ sub-problems:

\begin{subequations}\label{eq:sub-perturbed}
\begin{alignat}{3}
\forall S \in \textbf{S}: \notag \\
& \nabla_\lambda^S\mathcal{L}=g_S\left(X^S_{int}, (X^S_{ext})^k\right)=0\\
& \nabla_X^S\mathcal{L}=\nabla_X f_{obj}(X^S_{int}) + \notag \\ 
& \quad \nabla_X^T g_S\left(X^S_{int}, (X^S_{ext})^k\right) (\lambda_{int}^S, (\lambda^S_{ext})^k) -\mu^S_{int}= 0\\
& -\mathit{\mu^S_{int}}\odot\left(X^S_{int}-\underline{X^S_{int}}\right)+\epsilon=0\\
& \mathit{\mu_{int}^S}\geq0\\
&X\geq\underline{X}
\end{alignat}
\end{subequations}

\begin{algorithm}
	\caption{WLAV Combined T\&D ACSE} 
	\label{alg:framework}
	\KwIn{T\&D network models, measurements, weights, interconnection nodes}
	\KwOut{Transmission and distribution grid estimates}
	{\bf Read T\&D input files}
	
	{\bf Domain-based  Decomposition:} Obtain a set of subsystems $\{S\}$ by domain-based decomposition.
	
	{\bf Initialize} epoch $k=0$. 
	
	\For{epoch k=1 to N}{
	\For{each sub-system S in parallel}{
	\If {k=1} {then initialize coupling port variables $X_{ext}^{S, k}, \lambda_{ext}^{S,k}$}
	
	{\bf Solve state estimation sub-problem $S$} in (\ref{eq:sub-perturbed}) with PDIP algorithm (using Newton's method).
	
	{\bf Transfer solutions of $S$ to the supervisory layer.}
	}

	{\bf Supervisory Layer:}  Collect limited information for all sub-systems $\textbf{S}$ and calculate/update $X_{ext}^{S}$ based on (\ref{eq: T&D current coupling}, \ref{eq: T&D voltage coupling}, \ref{eq:supervisory layer}) to be used in $(k+1)^{th}$ epoch.
	
	{\bf Check convergence and break if converged}
	}
\end{algorithm}

In the described estimation algorithm (also see Algorithm \ref{alg:framework}), each individual entity solves the independent sub-problem $S$ iteratively through Newton's method to complete one epoch. After each iteration the external variables $X_{ext}$ and $\lambda_{ext}^S$ are communicated to the supervisory layer that solves a set of coupling equations (\ref{eq: T&D current coupling})-(\ref{eq: T&D voltage coupling}) to describe the Gauss step:
\begin{equation}
    f_{ext}(X^k, \lambda^k) = 0
    \label{eq:supervisory layer}
\end{equation}

\noindent For the $(k+1)^{th}$ epoch, external variables $X_{ext}$ calculated from the solution of $k^{th}$ epoch are used. Convergence is achieved when the norm of change in external variables  between two subsequent epochs $(\norm{X_{ext}^{k+1} - X_{ext}^k})$ is less than a predefined tolerance $\epsilon$.  The algorithm satisfies a key requirement (\ref{req:dim requirements in decomposition}) for the given domain-based decomposition, which is that the dimension of internal states within any sub-circuit is much larger than the dimension of external states within that sub-circuit. This condition is naturally satisfied in T\&D circuits.
\begin{alignat}{2}
    \text{dim}(X^S_{int}) \gg \text{dim}(X^S_{ext}) \quad \forall S \in \{1,....,|\textbf{S}|\}
    \label{req:dim requirements in decomposition}
\end{alignat}

\noindent Another nice-to-have property for such a framework is for individual entities to keep measurement data $(z_s)$ private without having to share it with external entities. This condition can be satisfied by applying the domain-based decomposition with zero-injection nodes as the boundary between different sub-networks.

%% file: Experiments.tex
\section{Experiments} \label{heading: Experiments}

\subsection{Problem Setup and Creation of Synthetic Measurements}
This section evaluates the performance of the developed combined T\&D ACSE algorithm. To generate synthetic measurements for the ACSE experiments, we first ran a set of combined T\&D power flow simulations following the methodology in \cite{Pandey_CombinedTD}. The output of these represented synthetic true states for the simulated T\&D configuration. We generated synthetic measurements by adding Gaussian noise to these \textit{true} states. With access to synthetic measurements, for transmission networks, we constructed positive sequence measurement models at all injection nodes following methodology in Section \ref{heading: Section III}. Additionally, we also constructed line flow measurement models from $P_{line}^P,Q_{line}^P$ measurements for some randomly selected flow measurements. For distribution network, we constructed three-phase measurement models for all load nodes (from $|V|^{\delta},P_{inj}^{\delta},Q_{inj}^{\delta}$). For the second group of experiments with bad-data, we created measurement samples with synthetic bad-data by adding large deviations to a few randomly selected measurements in each of the sub-networks. 

For all experiments, we evaluate the performance of the ACSE algorithm through use of root mean square error $RMSE$ metric, which is defined as follows:
$$RMSE = \sqrt{||x_{est}-x_{true}||_2^2/\text{dim}(x)}$$

\noindent where $x_{est}$ is a vector of estimated states in the T\&D network and $x_{true}$ is a vector of true states obtained from the combined T\&D power flow simulation. A lower value of RMSE suggests that the grid estimate is closer to its true magnitude. To run the distributed ACSE framework, we made use of Google cloud infrastructure with multiple compute instances. The details are:  \#cores: 16, Total RAM: 128 GB and CPUs: Intel Cascade Lake.


\subsection{Experiment 1: Combined T\&D ACSE (WLAV and WLSE) with Gaussian Noise}

For the first experiment, we ran the WLAV- and WLSE- ACSE algorithms on two combined T\&D networks. The first combined T\&D network consisted of a 10k node synthetic Texas transmission network (ACTIVSg10k) coupled with 7 realistic distribution feeders (see taxonomy feeders in \cite{taxonomy_feeders}). The second network consisted of a larger transmission network with 25k nodes (ACTIVSg25k) with an additional meshed urban distribution feeder.
The results from the ACSE runs are documented in Table \ref{tab:RMSE results}. As expected, with no presence of bad-data the performance of both WLSE and WLAV estimation algorithms is similar, with WLAV algorithm providing marginally better estimates.

\subsection{Experiment 2: Combined T\&D ACSE (WLAV and WLSE) with Bad-data}

In the second experiment, we added bad-data to a few randomly selected measurements ($<$15 in each sub-network) to the same T\&D networks that were evaluated in Experiment 1. We repeated the experiment in presence of bad-data for both combined T\&D networks and documented the results in Table \ref{tab:RMSE results}. The results show that the RMSE performance metric for the WLAV algorithm is lower than the WLSE algorithm. This implies that the WLAV estimates are closer to the true grid states than the WLSE estimates, hence validating the property of the WLAV algorithm to be robust against bad-data.


\begin{table}[ht]
\caption{RMSE metric comparison on combined T\&D networks}
\label{tab:RMSE results}
\begin{tabular}{ccc} 
 \hline
  \bf T\&D CASE & \begin{tabular}[c]{@{}c@{}} \bf Circuit-theoretic\\ \bf WLAV SE \end{tabular} & \begin{tabular}[c]{@{}c@{}} \bf Circuit-theoretic\\ \bf WLS SE \end{tabular} \\
 \hline
 \begin{tabular}{c}
      ACTIVSg10k + 7 D-nets$^1$
 \end{tabular} & 0.00465 & 0.00536\\ 
 \hline
 \begin{tabular}{c}
 ACTIVSg10k + 7 D-nets$^1$\\ 
 \color{red} (with bad-data) \end{tabular}  & 0.00556 & 0.01509  \\
 \hline
  \begin{tabular}{c}
      ACTIVSg25k + 8 D-nets$^2$
 \end{tabular} & 0.00848  & 0.00893\\ 
 \hline
 \begin{tabular}{c} 
 ACTIVSg25k + 8 D-nets$^2$\\
 \color{red} (with bad-data) \end{tabular}  & 0.00881 & 0.01335  \\ 
 \hline
\end{tabular}\\
\footnotesize{* D-net: Distribution network.\\
$^1$ Taxonomy feeders GC-12.47-1 and R1-12.47-3 represent the set of distribution feeders.\\
$^2$ In addition to the taxonomy feeders, a larger urban meshed distribution system Network\_Model\_Case1 is used.\\
* Both circuit-theoretic WLAV and WLS SEs are proposed within this paper. The difference is that WLAV-SE minimizes $||\eta||_1$ and is converted to (\ref{eq:problem_definition}) with $\eta^{+/-}$ as noise variables, while WLS SE minimizes $||\eta||_2^2$.}
\end{table}

\begin{figure}[htp]
    \centering
    \includegraphics[width=7.5cm]{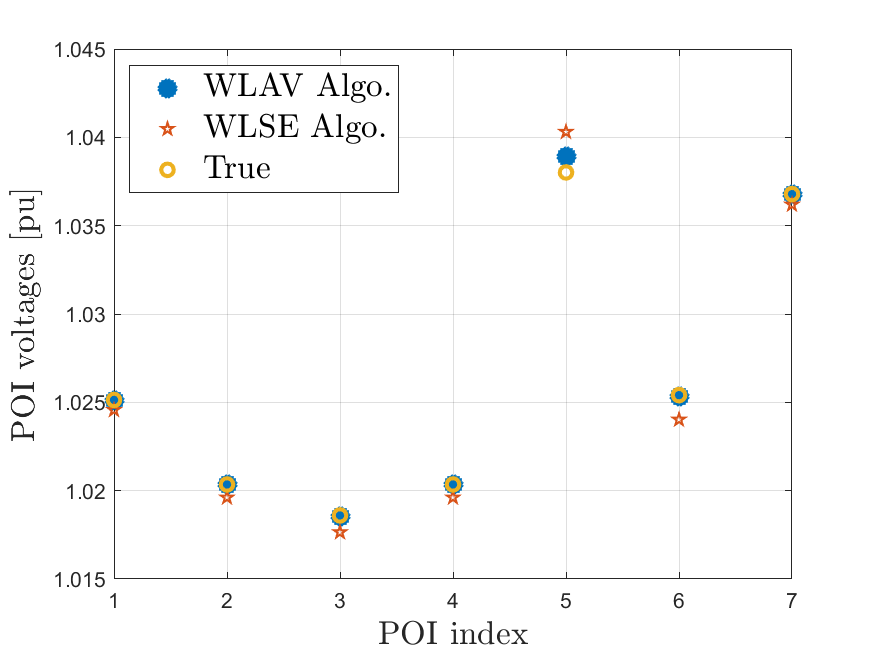}
    \caption{Voltage estimates at point of interconnection for combined T\&D model ACTIVSg10k + 7 D-nets.}
    \label{fig:poi_voltages}
\end{figure}

To further validate the efficacy of the combined T\&D SE algorithms in presence of bad-data, we plot the true (obtained from combined T\&D power flow) as well as estimated voltages at the various (7) point of interconnections (POIs) between T\&D sub-networks in Figure \ref{fig:poi_voltages}. For the estimated values, we plot voltages obtained from both WLSE and WLAV ACSE algorithms. We see that although both WLAV and WLSE provide good estimates for the POI voltages, WLAV algorithm provides more accurate estimates as is inferred from the distance between true and estimated values. This validates that the WLAV algorithm provides more accurate estimates under presence of bad-data as compared to WLSE algorithm.

\subsection{Experiment3: Scalability}

\begin{figure}[htp]
    \centering
    \includegraphics[width=6cm]{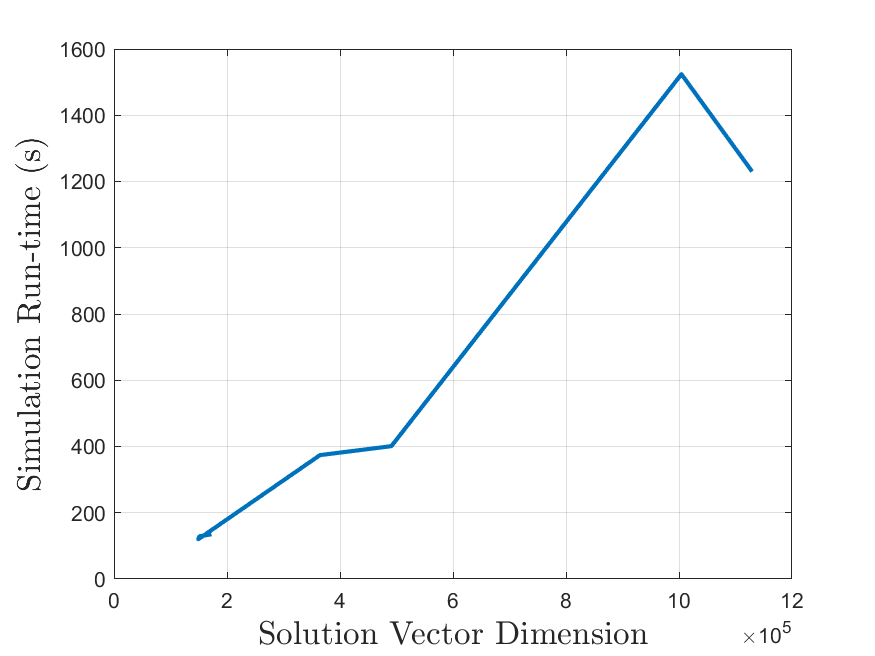}
    \caption{Combined T\&D WLAV ACSE algorithm scalability.}
    \label{fig:scalability}
\end{figure}

Finally, to demonstrate the scalability of the proposed algorithm, we ran combined T\&D ACSE simulations on a set of networks with solution vector dimension varying from a few thousands to a million+ and plotted the simulation time (in sec.) on the vertical axis in Figure \ref{fig:scalability}. We observed that the speed of the simulation is highly dependent on the size of the largest sub-network. For instance, the time for the million+ dimension system is found to be dependent on the run-time of the 70k+ node transmission network. Nonetheless, even with the prototype implementation of the tool, we demonstrated that the large T\&D networks can be solved within 15 min.

%% file: Conclusion.tex
This paper developed a novel circuit-theoretic approach for combined T\&D ACSE problem. The approach is distributed both in memory and compute resources to support the practical needs of the problem with its data stored across multiple utilities and sheer number of the solution variables. The  approach is able to obtain a truer or more physical solution by capturing the nodal relationships for all T\&D nodes in the network. Also, by formulating the distributed estimation problem with weighted least-absolute value objective, the approach can implicitly reject bad-data. In the results, we demonstrate these features of the approach through runs on a set of large T\&D networks both with and without bad-data.